\documentclass[]{spie}  

\newcommand{\hs}{\hspace{8pt}}
\newcommand{\meanF}{\ensuremath{\langle W \rangle}}
\newcommand{\meanM}{\ensuremath{\langle M \rangle}}
 
\usepackage{amsmath,amsfonts,amssymb}
\usepackage{graphicx}
\usepackage[colorlinks=true, allcolors=blue]{hyperref}

\title{Sex-Disaggregated Systematics in Canadian Time Allocation Committee Telescope Proposal Reviews}

\author[a]{Kristine Spekkens}
\author[b]{Nicholas Cofie}
\author[c]{Dennis R. Crabtree}

\affil[a]{Royal Military College of Canada and Queen's University}
\affil[b]{Queen's University}
\affil[c]{Herzberg Astronomy and Astrophysics, National Research Council of Canada}

\authorinfo{Email: kristine.spekkens@rmc.ca}

\begin{document} 
\maketitle

\begin{abstract}
Recent studies have shown that the proposal peer review processes employed by a variety of organizations to allocate astronomical telescope time produce outcomes that are systematically biased depending on whether proposal's principal investigator (PI) is a man or a woman. Using Canada-France-Hawaii Telescope (CFHT) and Gemini Observatory proposal statistics from Canada over 10 recent proposal cycles, we assess whether or not the mean proposal scores assigned by the National Research Council's (NRC's)  Canadian Time Allocation Committee (CanTAC) also correlate significantly with PI sex. Classical $t$-tests, bootstrap and jackknife replications show that proposals submitted by women were rated significantly worse than those submitted by men. We subdivide the data in order to investigate sex-disaggregated statistics in relation to PI career stage (faculty vs.\ non-faculty), telescope requested, scientific review panel, observing semester, and the PhD year of faculty PIs. Consistent with the bivariate results, a multivariate regression analysis controlling for other covariates confirmed that PI sex is the only significant predictor of proposal rating scores for the sample as a whole, although differences emerge for proposals submitted by faculty and non-faculty PIs. While further research is needed to explain our results, it is possible that implicit social cognition is at work. NRC and CanTAC have taken steps to mitigate this possibility by altering proposal author lists in order to conceal the PI's identity among co-investigators. We recommend that the impact of this measure on mitigating bias in future observing semesters be quantitatively assessed using statistical techniques such as those employed here.

\end{abstract}


\section{INTRODUCTION}
\label{sec:intro}  

Observing time on research-class astronomical telescopes is typically granted through peer review processes that aim to identify proposals with the highest scientific and technical merit and prioritize them for telescope access. With few (relatively recent) exceptions, the names of the principal investigator (PI) and co-investigators of a proposal are known to the individuals assessing its merit. Recent studies of the telescope proposal review mechanisms employed by a number of organizations suggest that, in addition to scientific and technical merit as intended, PI sex also influences proposal rankings. In a sex-disaggregated statistical study of Hubble Space Telescope (HST) data for Cycles 11--21, Reid \cite{Reid14} found that proposals submitted by men are more likely to be accepted than those submitted by women. A similar study conducted by Patat \cite{Patat16} finds that women are significantly less likely to be allocated observing time on European Southern Observatory (ESO) facilities than men. Lonsdale et al.\ \cite{Lonsdale16} found that the proposal review processes for facilities operated by the National Radio Astronomy Observatory (NRAO) also favour men over women, with the most significant differences stemming from the Atacama Large Millimeter Array (ALMA) Proposal Review Processes for Cycles 2-4. These sex-disaggregated statistics from telescope proposal reviews are part of a broader picture in which peer review outcomes for women are less favourable than those for men \cite{Wenneras97,Conley12,Moss12,West13,Caplar16,King16}, itself one facet of the complex issue of gender statistics in academia \cite{UN12}.



In this contribution, we examine recent proposal review statistics from the National Research Council's (NRC's) Canadian Time Allocation Committee (CanTAC) for access to the Canadian share of Canada-France-Hawaii Telescope (CFHT) and Gemini Observatory time, in order to assess whether or not sex-disaggregated mean proposal scores differ. We also investigate sex-disaggregated statistics in relation to PI career stage (faculty vs.\ non-faculty), the telescope requested, scientific review panel, observing semester, and the PhD year of faculty PIs. We present the available data in Section~\ref{sec:data}, and outline our statistical methods in Section~\ref{sec:methods}. We highlight some results from our statistical analysis in Section~\ref{sec:results}, and discuss our findings in Section~\ref{sec:discuss}.


\section{THE DATA: CANTAC PROPOSAL SCORES}
\label{sec:data}

CanTAC's main function is to  review proposals for access to Canada's share of CFHT and Gemini observing time\footnote{https://www.nrc-cnrc.gc.ca/observing/cantac.html} on the basis of scientific and technical merit. CanTAC is typically comprised of 12 academics with a range of scientific and technical expertise, and CanTAC members serve three-year terms reviewing proposals on either the Extragalactic or Galactic panel. Reports from up to three anonymous external referees as well as a technical review by observatory staff are solicited for each proposal by NRC, and the resulting feedback is provided to CanTAC along with the proposal itself to inform their review. In a given semester, the Extragalactic or Galactic panel ranks each proposal that it is tasked to review by assigning a linear-rank score from 1 -- 6, where a {\it lower} number implies {\it higher} scientific and technical merit, and is therefore better. Normalized scores from the Extragalactic and Galactic panels are then interleaved into a single ranked list that, in combination with instrument availability, weather conditions and constraints from partner countries, ultimately dictates which proposals are granted observing time. CanTAC's normalized proposal scores are therefore strong predictors of access to Canada's share of CFHT and Gemini time.

\begin{table}[t]
\caption{Descriptive Statistics}
\label{tab:descriptive}
\begin{center}
\begin{tabular}{l|ccc|ccc|ccc}
\hline
{\bf Attribute} & \multicolumn{3}{|c|}{\bf All PIs} & \multicolumn{3}{|c|}{\bf Faculty PIs} & \multicolumn{3}{|c}{\bf Non-faculty PIs} \\
                     & {\bf All}  & {\bf Men} & {\bf Women}  & {\bf All}  & {\bf Men} & {\bf Women}  & {\bf All}  & {\bf Men} & {\bf Women}  \\
          & {\bf (\%) }    &  {\bf (\%) }      & {\bf (\%)  }   & {\bf (\%) }    &  {\bf (\%) }      & {\bf (\%)  }   & {\bf (\%) }    &  {\bf (\%) }      & {\bf (\%)  }       \\
  { \bf (1)}   & {\bf (2) }    &  {\bf (3) }      & {\bf (4)  }   & {\bf (5) }    &  {\bf (6) }      & {\bf (7)  }   & {\bf (8) }    &  {\bf (9) }      & {\bf (10)  }       \\

\hline
  {\bf All}       & 100.0   &  72.7    &  27.3   & 100.0 & 78.5 &      21.5 & 100.0 & 69.8 & 39.2    \\
 {\bf Telescope} &       &  &  &       &  &  &       &  & \\
 \hs CFHT & 41.0 & 41.8 & 39.0 &  40.0 & 42.6 & 34.1 & 37.7 & 36.4 & 40.7\\
 \hs Gemini & 59.0 & 58.2 & 61.0 & 60.0 & 58.4 & 65.9 & 62.3 & 63.6 & 59.3\\
 {\bf Panel} &       &  &  &       &  &  &       &  & \\
  \hs Extragalactic & 43.7 &  42.7 & 46.3 &   57.6    & 56.5 & 61.4 &   36.4     & 34.0 & 42.1\\

 \hs Galactic & 56.3 &  57.3 &  53.7&   42.4 & 43.5 & 38.6 &  63.6    & 66.0  & 57.9\\

  {\bf Semester} & & & &      & & &      & &\\
 \hs 2012A & 7.7 & 7.5 & 8.3 & 11.2 & 11.8 & 9.1 & 6.7 & 5.8 & 8.6\\
 \hs 2012B & 11.7 & 11.6 &  12.2  & 10.7 & 11.8 & 6.8 & 13.2 & 13.0 & 13.6\\
 \hs 2013A & 8.8 & 8.6 & 9.3 & 11.7 & 11.2& 13.6 & 7.1 & 7.1 & 7.2\\
 \hs 2013B & 10.0 & 11.4 & 6.4 & 8.3 & 8.7 & 6.8 & 10.8 & 12.7 & 6.4\\
 \hs 2014A & 8.7 & 9.0 & 7.8 & 11.2 & 11.8 & 9.1 & 7.5 & 7.7 & 7.1\\
 \hs 2014B & 9.8 & 9.5 & 10.7 & 9.3 & 9.3 & 9.1 & 9.9 & 9.6 & 10.7\\
 \hs 2015A & 17.0 & 16.5 & 18.5 & 13.7 & 13.0 & 15.9 & 18.1 & 17.0 & 20.7\\
 \hs 2015B & 7.6 & 8.4 & 5.4 & 5.9 & 6.8 & 2.3 & 7.3 & 8.0 & 5.7\\
\hs  2016A & 11.5 & 11.5 & 11.2 & 11.7 & 10.6 & 15.9 & 11.6 & 12.3 & 10.0\\
\hs 2016B & 7.2 & 6.0 & 10.2 & 6.3 & 5.0 & 11.4 & 7.8 & 6.8 & 10.0\\
{\bf PhD Year}$^a$ &      & & &      & & &      & &\\
 \hs $\leq$ 1989 &   ...   & ...& ...&  15.1    & 19.2 & 0.0 &   ...   & ...& ...\\
 \hs 1990 -- 1999 &   ...   & ...& ...&   53.7     & 51.6 &  61.4 & ...& ...& ...\\
 \hs $\geq 2000$ &    ...  & ...& ...&    31.2    & 29.2 &  38.6  & ...& ...& ...\\
\hline
{\bf N} & {\bf 751} & {\bf 546} & {\bf 205} & {\bf 205}  & {\bf 161} & {\bf 44} & {\bf 464} & {\bf 324} & {\bf 140}\\
\hline
\end{tabular}
\end{center}
$a$: only available for faculty PIs.
\end{table} 

	The current study considers all $N=751$ proposals requesting either CFHT or Gemini time that were reviewed by CanTAC in 10 recent semesters, from 2012A -- 2016B inclusively. The information made available to us that was collected from PIs by NRC during the proposal submission and review process includes the semester in which the proposal was submitted for review, the telescope requested, and the scientific panel (Extragalactic vs.\ Galactic) that reviewed it.  
	
	In 2016, additional PI information was reverse-engineered from the proposals by NRC on the basis of the PI names (to which we did not have access during this study) and also provided to us. For $N=669$ proposals, it was determined whether or not the PI was on faculty at an academic institution at the time the proposal was submitted, and if so, the year in which they obtained their PhD. A binary determination of the sex (man or woman) of the PI was also made for all $N=751$ proposals. For conciseness in what follows, we use the term ``men" and ``women" to describe PIs labelled as such during the course of this determination, and ``PI sex" to describe the label itself. We note that self-declarations encompassing a wide range of gender identities are clearly preferable to the post-facto binary sex labels adopted here. In the absence of this information, however, we proceed with the available data in the hopes that it will provide some insight into the broader issue of gender statistics in scientific reviews \cite{UN12}.


\begin{table}[t]
\caption{t-tests Examining Differences in Mean Proposal Scores}
\label{tab:ttests}
\begin{center}
\begin{tabular}{l|cccc|cccc|cccc}
\hline
 {\bf Attribute}       & \multicolumn{4}{|c|}{\bf All PIs} & \multicolumn{4}{|c|}{\bf Faculty PIs} & \multicolumn{4}{|c}{\bf Non-faculty PIs} \\
& {\bf \meanF}  & {\bf \meanM} & {\bf $t$}  & {\bf $p$}  & {\bf \meanF} & {\bf \meanM}  & {\bf $t$}  & {\bf $p$} &   {\bf \meanF}  & {\bf \meanM} & {\bf $t$}  & {\bf $p$}  \\
  { \bf (1)}   & {\bf (2) }    &  {\bf (3) }      & {\bf (4)  }   & {\bf (5) }    &  {\bf (6) }      & {\bf (7)  }   & {\bf (8) }    &  {\bf (9) }      & {\bf (10)  }  &    {\bf (11)  }  &   {\bf (12)  }  & {\bf (13)  }  \\
\hline
  {\bf All}      & 3.60  &  3.45    &  3.40  & 0.000 & 3.45 & 3.36 & 0.92 & 0.179 & 3.67 & 3.46 & 2.23 & 0.006     \\
   {\bf Telescope} &       &  &  &       &  &  &       &  & & & &  \\
 \hs CFHT & 3.59 & 3.46 & 1.76 & 0.039 & 3.50 & 3.36 & 0.86 & 0.192 & 3.56 & 3.49 & 0.74 & 0.229 \\
 \hs Gemini & 3.61 & 3.44 & 2.93 & 0.002 & 3.42 & 3.78 & 0.46 & 0.322 & 3.64 & 3.45 & 2.59 & 0.005 \\
   {\bf Panel} &       &  &  &       &  &  &       &  & & & &  \\
 \hs Extragalactic & 3.66 & 3.43 & 3.36 & 0.000 & 3.42 & 3.30 & 1.07 & 0.143 & 3.51 & 3.75 & 2.63 & 0.005 \\
  \hs Galactic & 3.47 & 3.35 & 1.47 & 0.072 & 3.49 & 3.46 & 0.25 & 0.401 & 3.50 & 3.44 & 0.79 & 0.215 \\
 {\bf Semester} & & & &      & & &      & & & & &\\
 \hs 2012A & 3.62 & 3.45 & 1.09 & 0.140 & 3.66 &  3.39 & 0.98 & 0.167 & 3.57 & 3.52 & 0.23 & 0.408\\
 \hs 2012B & 3.52 & 3.50 & 0.14 & 0.445 & 3.35 & 3.68 & -0.86 & 0.200 & 3.54 & 3.44 & 0.60 & 0.277 \\
 \hs 2013A & 3.69 & 3.44 & 1.70 & 0.047 & 3.57 & 3.31 & 0.94 & 0.178 & 3.69 & 3.50 & 0.87 & 0.196 \\
 \hs 2013B & 3.52 & 3.50 & 0.11 & 0.456 & 3.43 & 3.26 & 0.52 & 0.304 & 3.54 & 3.53 & 0.08 & 0.467 \\
 \hs 2014A & 3.91 & 3.36 & 3.67 & 0.000 & 3.82 & 3.33 & 2.04 & 0.027 & 3.91 & 3.34 & 2.52 & 0.008 \\
 \hs 2014B & 3.55 & 3.44 & 0.72 & 0.236 & 3.43 & 3.24 & 0.57 & 0.289 & 3.59 & 3.49 & 0.54 & 0.296\\
 \hs 2015A & 3.46 & 3.48 & -0.21 & 0.394 & 3.20 & 3.38 & -1.27 & 0.108 & 3.51 & 3.49 & 0.19 & 0.428\\
 \hs 2015B & 3.51 & 3.47 & 0.27 & 0.394 & 3.91 & 3.22 & ... & ... & 3.32 & 3.49 & -1.00 & 0.154 \\
\hs  2016A & 3.64 & 3.46 & 1.48 & 0.071 & 3.52 & 3.40 & 0.49 & 0.316 & 3.59 & 3.45 & 0.98 & 0.166\\
\hs 2016B & 3.73 & 3.33 & 2.67 & 0.005 & 3.08 & 3.33 & -1.07 & 0.143 & 3.88 & 3.56 & 2.91 & 0.003\\
 {\bf PhD Year} &      & & &      & & &      & & & & &\\
 \hs $\leq$1989 &   ...   & ...& ...&  ...   & ...  & 3.59 & ... & ... &  ...   & ...& ...& ...\\
 \hs 1990 -- 1999 &   ...   & ...& ...&   ... & 3.47   & 3.17  & 2.97 & 0.002 &  ... & ...& ...& ...\\
 \hs $\geq 2000$ &    ...  & ...& ...&   ... &  3.42  & 3.58 & -1.16 & 0.125 & ...  & ...& ...& ...\\
\hline
\end{tabular}
\end{center}
Cols.\ (2), (6), (10): mean proposal scores \meanF\ for women in the sample as a whole, in the faculty PI sample, and in the non-faculty PI sample, respectively, where a {\it lower} number implies {\it higher} scientific and technical merit, and is therefore better. Cols.\ (3), (7), (11): mean proposal scores \meanM\ for men in the sample as a whole, in the faculty PI sample, and in the non-faculty PI sample, respectively. Cols.\ (4), (8), (12): t-score $t$ against the null hypothesis that $\meanF = \meanM$. Cols.\ (5), (9), (13): one-tailed p-value $p$ corresponding to $t$.
\end{table} 

\begin{table}[ht]
\caption{Bivariate regression results}
\label{tab:bivariate} 
\begin{center} 
\begin{tabular}{lcc}
\hline
{\bf Sample} & {\bf $\beta$} & {\bf $p$} \\
{\bf (1)} & {\bf (2)} & {\bf (3)} \\
\hline
All PIs & -0.15 & 0.001\\
Faculty PIs & -0.08 & 0.357\\
Non-faculty PIs & -0.14 & 0.013\\
\hline
\end{tabular}\\
Col. (2):  Unstandardized regression coefficient $\beta$ for a\\ 
model in which PI sex is the only predictor of proposal\\
grades using the mean proposal scores for women as the\\
reference category. Col. (3): p-value $p$ corresponding to $\beta$. 
\end{center}
\vspace{10pt}
\end{table}

Descriptive statistics are given in Table~\ref{tab:descriptive} for the sample of proposals as a whole, as well as for subsamples including only faculty and non-faculty PIs, respectively.  Men submitted $72.7\%$ of all CFHT and Gemini proposals received by CanTAC from 2012A -- 2016B compared to $27.3\%$ submitted by women, while proportionately more non-faculty proposals were submitted by women ($39.2\%$). Across all subsamples and irrespective of PI sex, $\sim40\%$ of proposals requested CFHT time compared to $\sim60\%$ for Gemini. A similar fraction of proposals ($\sim45\%$) were reviewed by the Extragalactic panel for the sample as a whole, although a higher fraction proposals submitted by non-faculty PIs ($\sim65\%$) were reviewed by the Galactic panel. There is scatter in the number of proposals submitted each semester, with the fraction ranging from as low as $7.2\%$ in 2016B to as high as $17.0\%$ in 2015A across all proposals; however, these fractions are similar among men and women in both the faculty and non-faculty samples. We quantify this observation using Chi-squared tests to search for statistically significant associations between PI sex and telescope requested, the scientific panel that carried out the review, the observing semester, and (for the faculty PI subsample) the year in which the PhD was awarded; we find no such associations in the data.

\section{STATISTICAL METHODS}
\label{sec:methods}

We quantify sex-disaggregated systematics in the data described in Section~\ref{sec:data} by comparing the mean proposal scores for men (\meanM ) and women (\meanF ). We assess the statistical significance of mean score differences using classical t-tests against the null hypothesis that $\meanF = \meanM$. The t-scores $t$ and corresponding one-tailed p-values $p$ for a variety of data subsamples are given in Table~\ref{tab:ttests}. 

\begin{table}[t]
\caption{Multivariate regression results}
\label{tab:multivariate}
\begin{center}
\begin{tabular}{l|cc|cc|cc}
\hline
{\bf Attribute}      & \multicolumn{2}{|c|}{\bf All PIs} & \multicolumn{2}{|c|}{\bf Faculty PIs} & \multicolumn{2}{|c}{\bf Non-faculty PIs} \\
&  {\bf $\beta$}  & {\bf $p$}  & {\bf $\beta$} & {\bf $p$}  & {\bf $\beta$}  & {\bf $p$}  \\
  { \bf (1)}   & {\bf (2) }    &  {\bf (3) }      & {\bf (4)  }   & {\bf (5) }    &  {\bf (6) }      & {\bf (7)  }    \\
\hline 
  {\bf All (women)}      & -0.15  &  0.001   & -0.12 & 0.190  & -0.13 & 0.025     \\
   {\bf Telescope (CFHT) } &       &  &  &       &  &  \\
 \hs Gemini & -0.01 & 0.854 & 0.01 & 0.914 & -0.02 & 0.732 \\
    {\bf Panel (Extragalactic) } &       &  &  &       &  &  \\
 \hs Galactic & -0.00 & 0.966 & 0.13 & 0.077 & -0.13 & 0.013 \\
 {\bf Semester (2012A) } & & & &      & &  \\
 \hs 2012B & 0.01 & 0.886 & 0.19 & 0.223 & -0.07 & 0.558 \\
 \hs 2013A & 0.01 & 0.880 & -0.04 & 0.775 & 0.02 & 0.858 \\
 \hs 2013B & 0.02 & 0.828 & -0.12 & 0.479 & 0.02 & 0.869 \\
 \hs 2014A & 0.00 & 0.970 & 0.01 & 0.943 & -0.02 & 0.869 \\
 \hs 2014B & -0.03 & 0.781 & -0.16 & 0.317 & 0.00 & 0.993\\
 \hs 2015A &  -0.02 & 0.824 & -0.10 & 0.494 & -0.04 & 0.761\\
 \hs 2015B & 0.00 & 0.961 & -0.15 & 0.428 & -0.06 & 0.652\\
\hs  2016A & 0.01 & 0.895 & 0.01 & 0.935 & -0.03 & 0.797\\
\hs 2016B & -0.03 & 0.790 & -0.22 & 0.220 & 0.01 & 0.935\\
 {\bf PhD Year ($\leq$1989)} &  ...    & ... & -0.01 & 0.084     & ... & ...\\
\hline
\end{tabular}
\end{center}
Col.\ (1): Covariate (reference value is bolded and given in parentheses). Cols.\ (2), (4), (6): unstandardized multivariate regression coefficient $\beta$ for a model that assesses the effect of PI sex on proposal grades by controlling for the other covariates. Col. (3): p-value $p$ corresponding to $\beta$. 
\end{table}

We also performed bivariate and multivariate regressions to assess the predictive power of PI sex in relation to mean proposal scores. Our bivariate regressions assess PI sex as the only predictor of mean proposal score, using the mean proposal scores for women as the reference category. Our multivariate regressions assess the effect of PI sex on proposal scores while controlling for other potential covariates: telescope, panel, semester, and (for the faculty PI subsample) the PhD year. Unstandardized regression coefficients $\beta$ and corresponding p-values $p$ against the null hypothesis that $\beta=0$ for the bivariate and multivariate regressions performed are given in Tables~\ref{tab:bivariate}~and~\ref{tab:multivariate}, respectively.

We consider a t-test or regression result to be statistically significant if $p<0.05$, and we verify these results using bootstrap, Jacknife, and stratified random sampling in order to replicate the original estimates.  We replicate all results where $p<0.05$ with these techniques, suggesting that our findings are robust. We present the highlights from the t-tests and regression analyses in Section~\ref{sec:results} below.



\section{RESULTS}
\label{sec:results}

Figure~\ref{fig:alldata} illustrates the mean proposal scores for men and women in the sample as a whole (left), the subsample of faculty PIs (middle), and the subsample of non-faculty PIs (right), along with $p$ for the corresponding t-test (from Table~\ref{tab:ttests}) and bivariate regression analysis (from Table~\ref{tab:bivariate}); asterisks denote statistically significant differences. 

  \begin{figure} [t]
   \begin{center}
   \begin{tabular}{c} 
   \includegraphics[height=7cm]{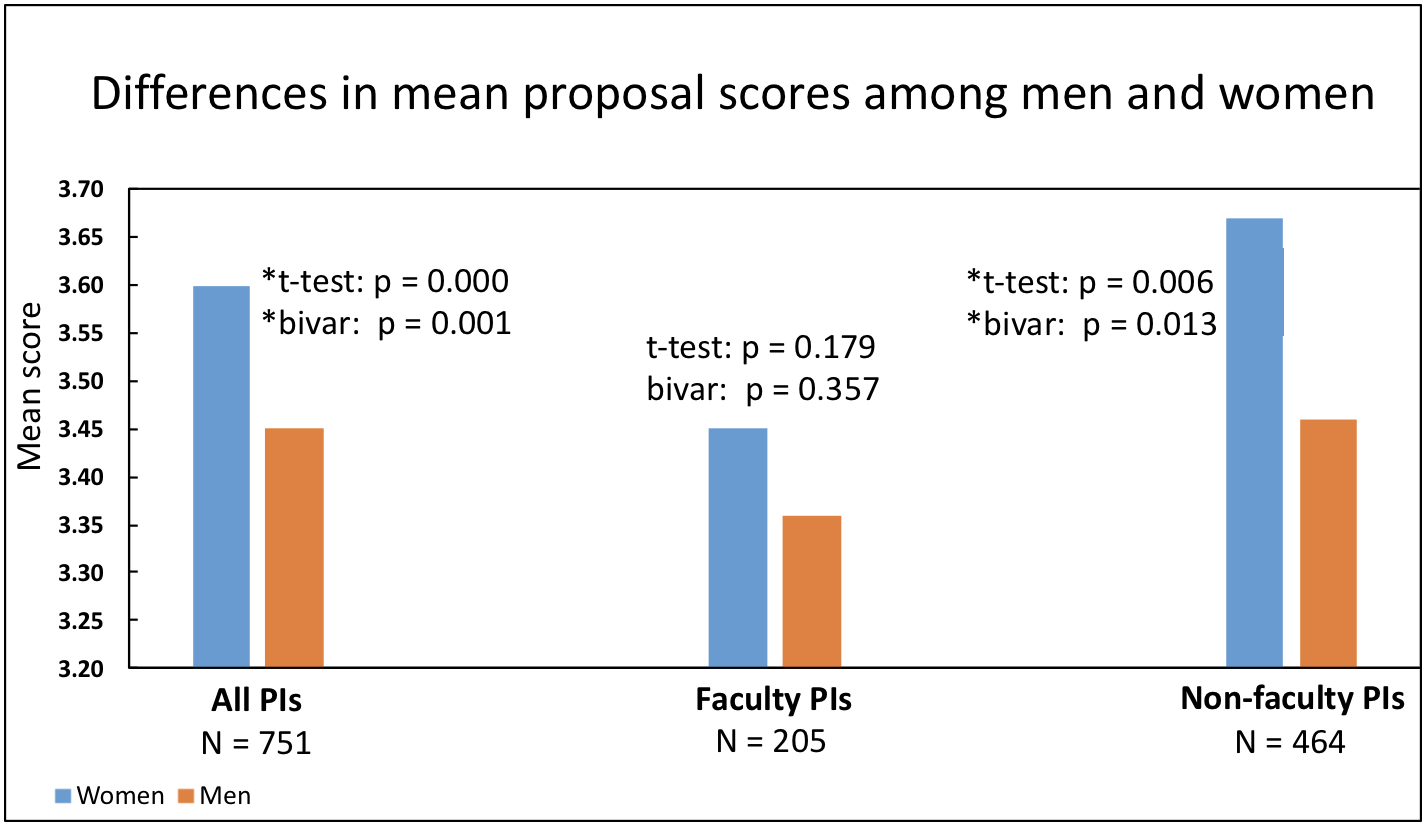}
	\end{tabular}
	\end{center}
   \caption[Differences in mean scores] 
   { \label{fig:alldata} 
Mean proposal scores for men (orange) and women (blue) for the sample as a whole (left), the faculty PI subsample (middle), and the non-faculty PI subsample (right). A {\it lower} number implies {\it higher} scientific and technical merit, and is therefore better. Values of $p$ from t-tests (Table~\ref{tab:ttests}) and bivariate regression analyses (Table~\ref{tab:bivariate}) are also given; asterisks denote statistical significance. {\bf The mean proposal scores for men and women in the sample as a whole differ significantly.}
}
   \end{figure} 
   
We find that the mean proposal scores for the sample as a whole differ significantly for men and women (Figure~\ref{fig:alldata} left), in the sense that men are more likely to obtain better proposal scores than women. Examining results for the faculty and non-faculty subsamples, we find statistically significant differences in mean proposal scores between men and women in the non-faculty PI subsample  (Figure~\ref{fig:alldata} right) but not in the faculty PI subsample (Figure~\ref{fig:alldata} middle). To further explore this result, we perform t-tests on sets of $N=205$ (the number of faculty PIs) proposals drawn from the larger non-faculty PI subsample using a stratified random sampling technique. The mean proposal scores for men and women  in these smaller subsets of the non-faculty subsample are significantly different; this suggests that the lack of a significant difference between men and women in the faculty PI subsample is not simply a result of the smaller subsample size, but instead implies a different underlying trend. 

\begin{figure} [ht]
   \begin{center}
   \begin{tabular}{c} 
   \includegraphics[height=7cm]{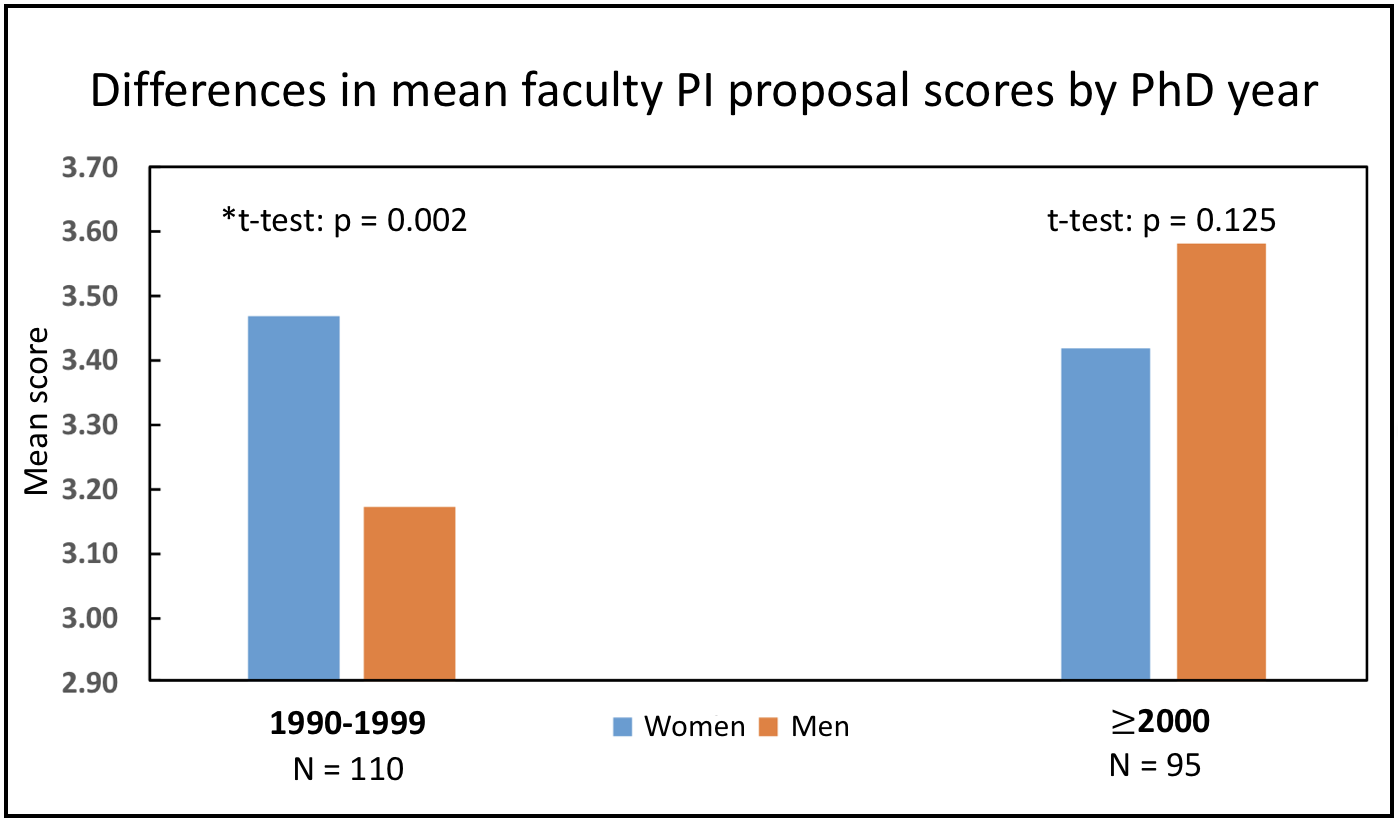}
	\end{tabular}
	\end{center}
   \caption[Differences in mean scores by PhD year] 
   { \label{fig:phdyear} 
Same as in Figure~\ref{fig:alldata}, but for faculty PIs with PhDs awarded between 1990 -- 1999 (left) and later than 2000 (right; there are no women awarded PhDs earlier than 1990 in our sample). {\bf Women granted PhDs between 1990 -- 1999 have significantly worse mean proposal scores than men awarded PhDs in the same time period.}
}
   \end{figure}

Figure~\ref{fig:phdyear} illustrates the one subset of the faculty PI subsample in which mean proposal scores do differ significantly according to our t-tests: women awarded PhDs between 1990 -- 1999 have significantly worse mean proposal scores than men awarded PhDs in the same time period. There are hints that this trend may reverse for faculty with PhDs awarded later than 2000, where $\meanF < \meanM$ (Figure~\ref{fig:phdyear}, right); however, that difference is not significant.  Indeed, $\meanF > \meanM$ for the majority of the faculty PI subsets that we examined (Table~\ref{tab:ttests}), Table~\ref{tab:multivariate} shows that we find no significant predictor of faculty PI proposal scores among the covariates explored. In what follows below, we highlight the statistically significant trends that we find in the sample as a whole and the subsample of non-faculty PIs. 

To better visualize the sex-dissagregated systematics in subsets of the data defined by the telescope requested, the scientific panel that reviewed the proposal and the observing semester in which the proposal was submitted for review, we plot the cumulative absolute deviation of the proposal scores $x_{M,i}$ for men relative to the mean score $\langle x \rangle$ of proposals submitted by men and women combined:
\begin{equation}
\label{eq:cum}
C_M = \sum^i (\langle x \rangle - x_{M,i}) \,\,\,,
\end{equation}
where $i$ sums over all proposals submitted by men. A value $C_M>0$ therefore indicates that men obtained better (= lower) proposal scores overall than women, while $C_M<0$ indicates that women obtained better scores. We also plot the corresponding cumulative absolute deviations for women in the same data subset, defined simply as $C_W = - C_M$.

\begin{figure} [t]
   \begin{center}
   \begin{tabular}{c} 
   \includegraphics[height=7cm]{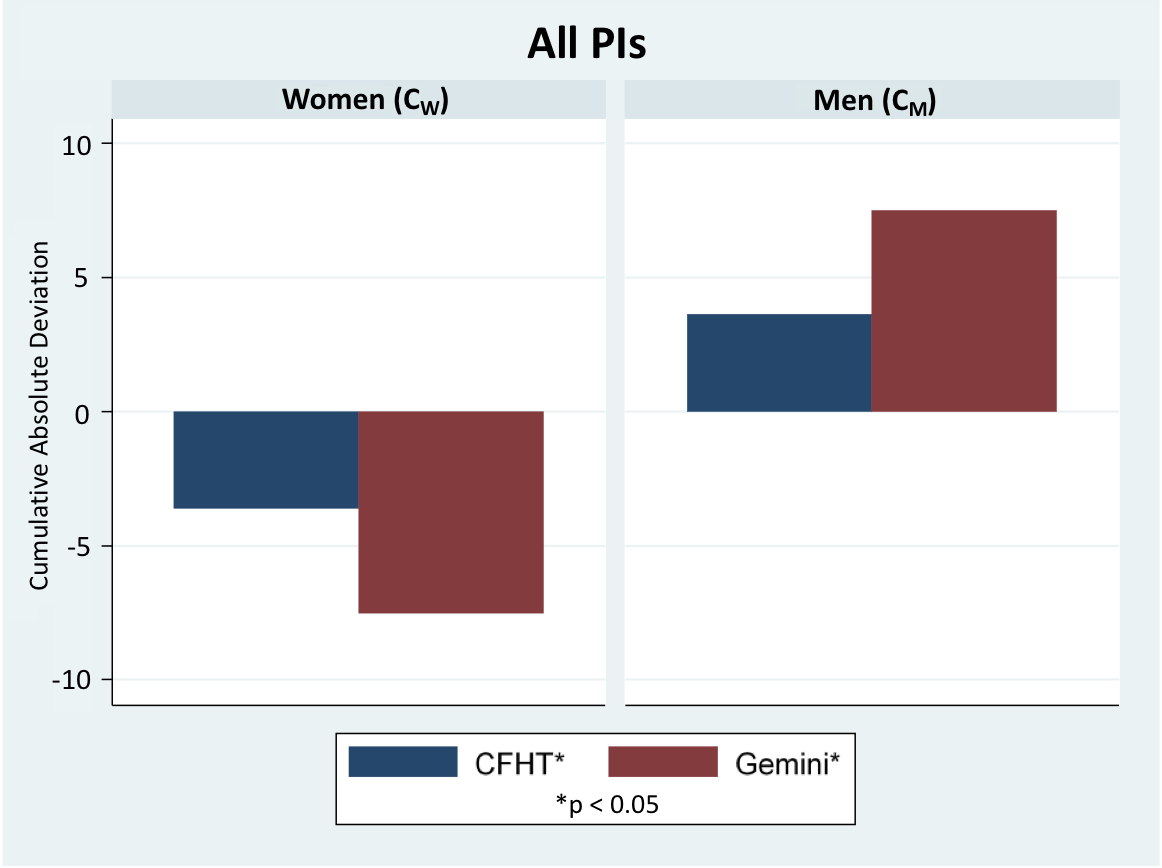}\\
    \includegraphics[height=7cm]{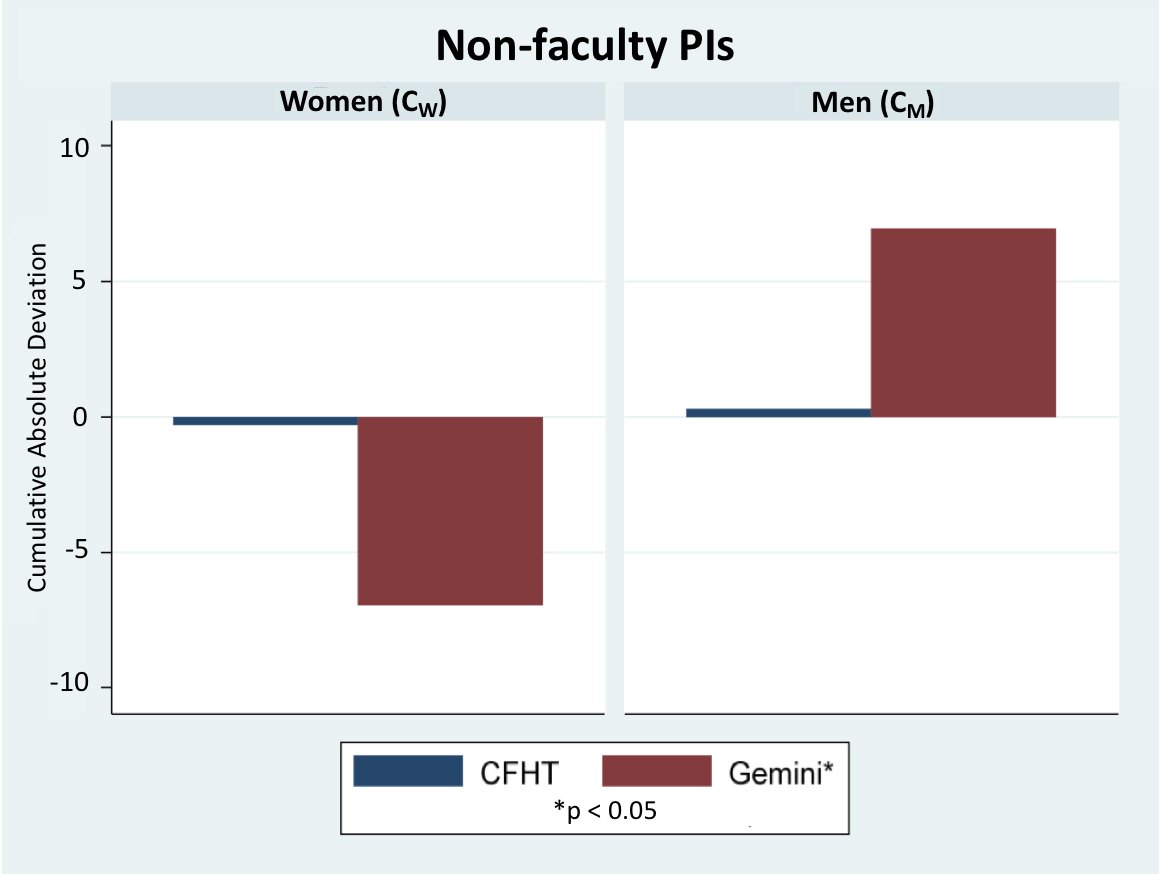}
	\end{tabular}
	\end{center}
   \caption[Differences in mean scores by telescope] 
   { \label{fig:telescopes} 
Cumulative absolute deviations (Equation~\ref{eq:cum}) in scores for women ($C_F$; left) and men ($C_M$; right) who submitted CFHT (blue) and Gemini (red) proposals, for the sample as a whole (top) and the subsample of non-faculty PIs (bottom). A value $C_M>0$ indicates that men obtained better (= lower) proposal scores overall than women, while $C_M<0$ indicates that women obtained better scores. Asterisks indicate statistically significant mean proposal score differences according to the t-tests reported in Table~\ref{tab:ttests}. {\bf Telescope access to both CFHT and Gemini differs significantly between men and women in the sample as a whole, as does Gemini access for non-faculty PIs.} 
}
   \end{figure}

\begin{figure} [t]
   \begin{center}
   \begin{tabular}{c} 
   \includegraphics[height=7cm]{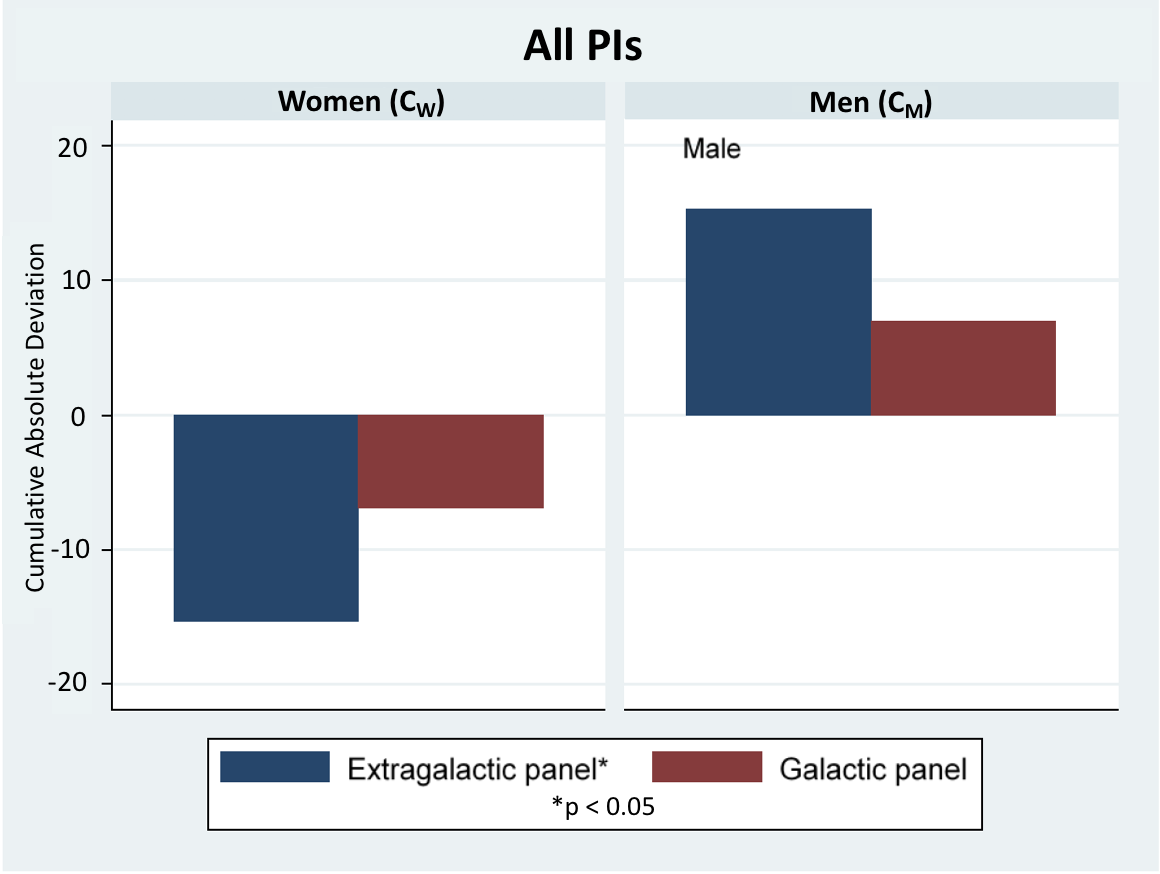}\\
   \includegraphics[height=7cm]{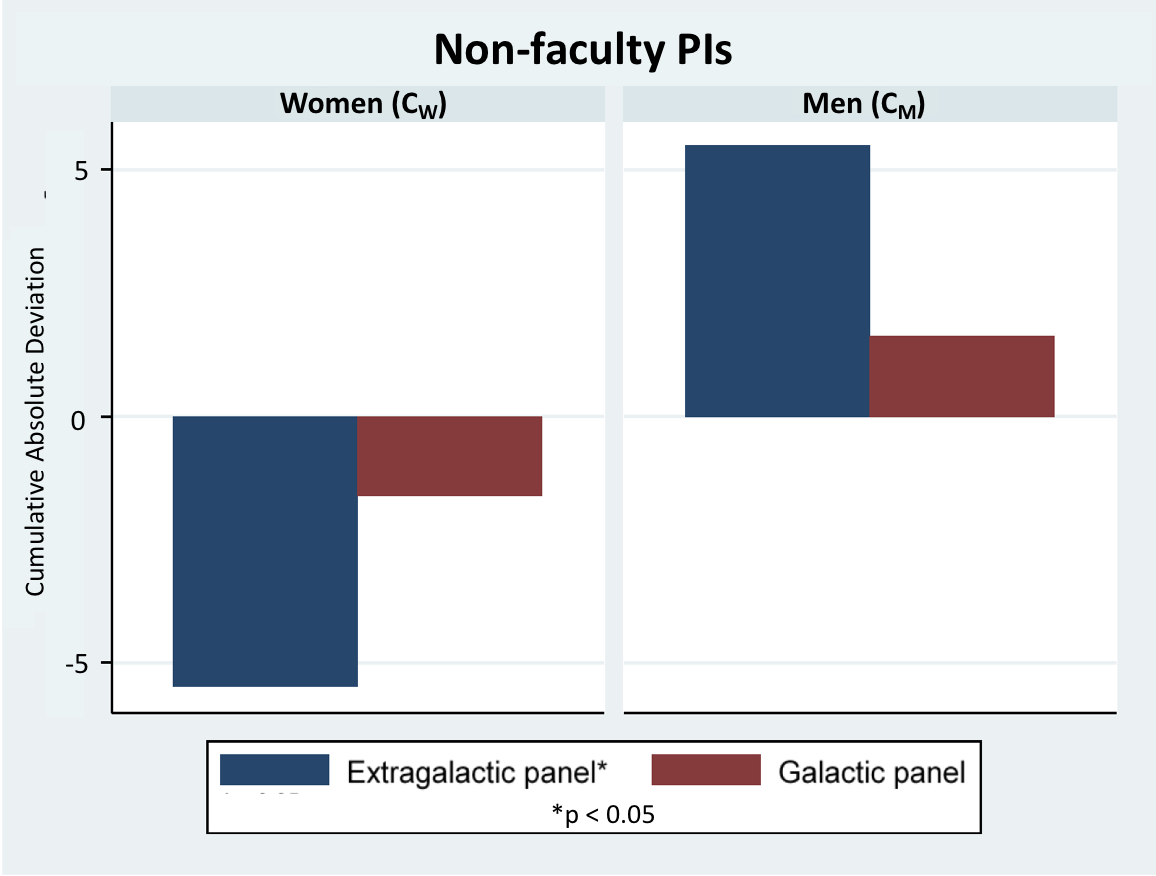}\\
	\end{tabular}
	\end{center}
   \caption[Differences in mean scores by panel] 
   { \label{fig:panel} 
Same as in Figure~\ref{fig:telescopes}, but for proposals evaluated by the Extragalactic panel (blue) and the Galactic panel (red).  {\bf Proposal evaluation by the Extragalactic panel differs significantly between men and women in the sample as a whole and in the subsample of non-faculty PIs.}
}
   \end{figure}

\begin{figure} [t]
   \begin{center}
   \begin{tabular}{c} 
   \includegraphics[height=7cm]{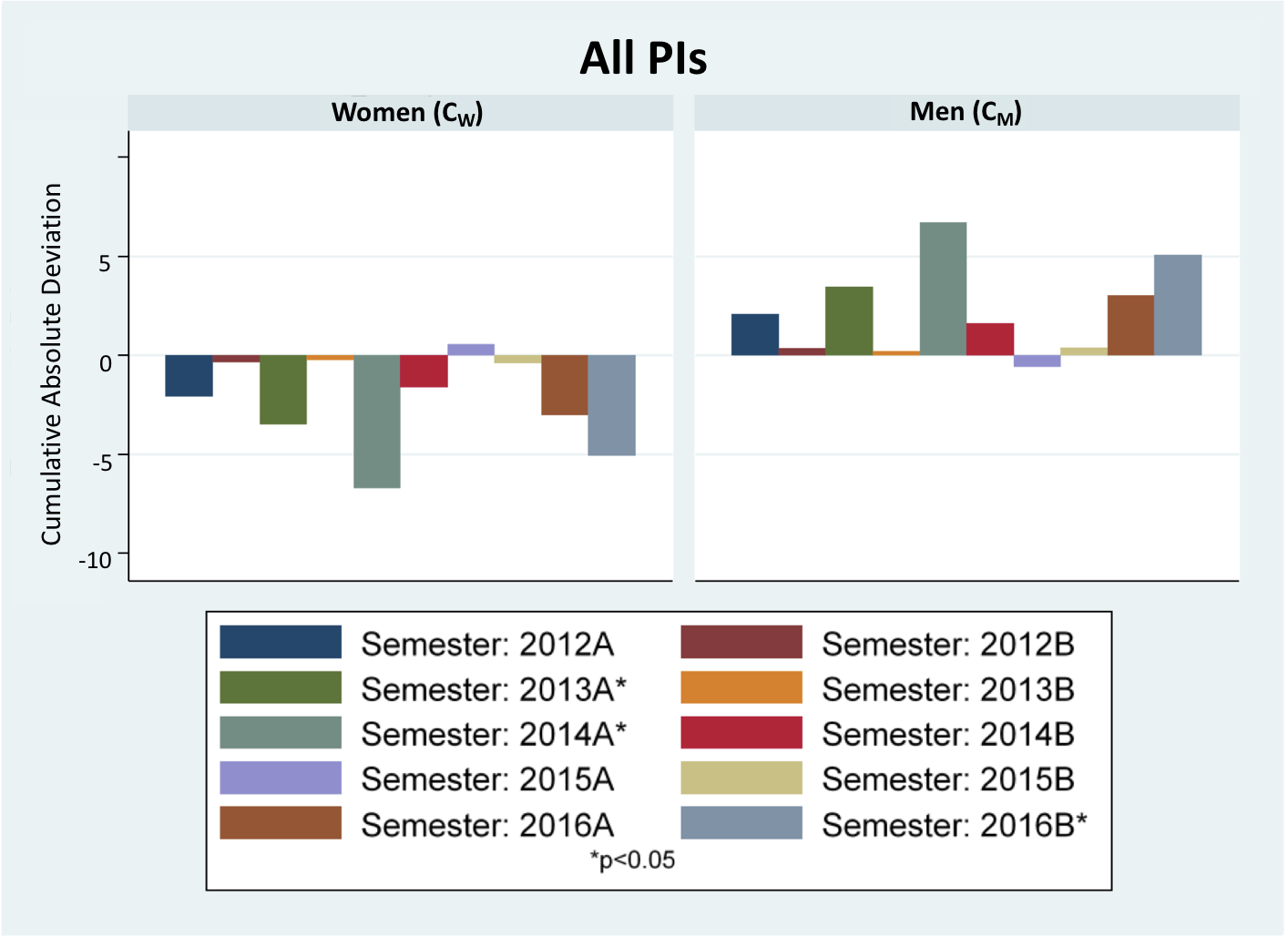}\\
   \includegraphics[height=7cm]{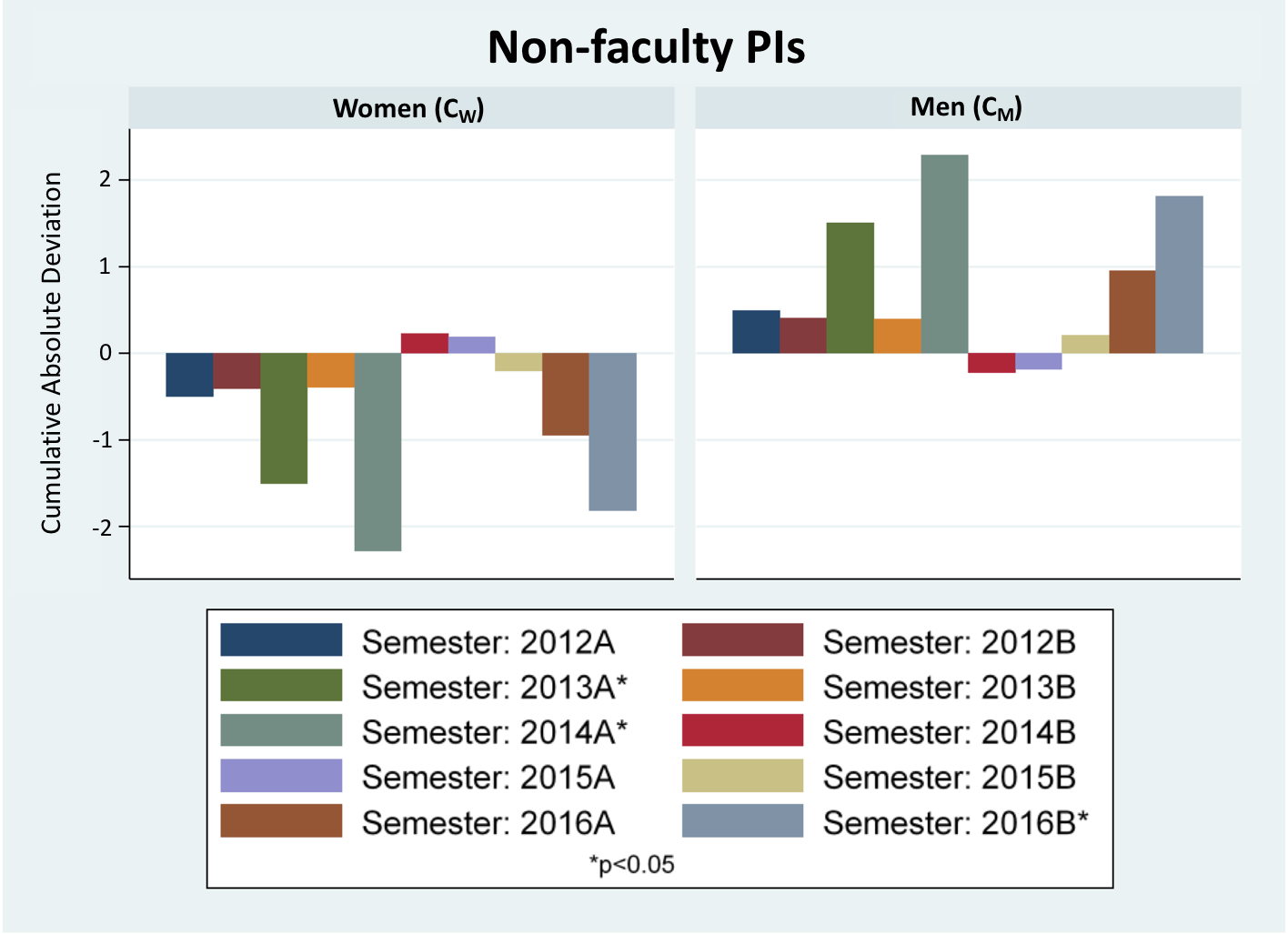}\\
	\end{tabular}
	\end{center}
   \caption[Differences in mean scores by semester] 
   { \label{fig:semester} 
Same as in Figure~\ref{fig:telescopes}, but for proposals submitted in different observing semesters.  {\bf Women in the sample as a whole and in the non-faculty PI subsample have worse mean scores than men in almost every semester, and significantly so in 2013A, 2014A and 2016B.} 
}
   \end{figure}

Figures~\ref{fig:telescopes},~\ref{fig:panel}~and~\ref{fig:semester} show cumulative absolute deviations as a function of telescope, panel and observing semester, respectively. In each figure, the top panel shows results for the sample as a whole, and the bottom panel shows results for the subsample of non-faculty PIs. Asterisks indicate statistically significant mean proposal score differences according to the t-tests reported in Table~\ref{tab:ttests}. We find that access to Gemini and proposals reviewed by the Extragalactic panel differ significantly in favor of men in both the sample as a whole and the non-faculty PI subsample. Figure~\ref{fig:semester} shows that women obtained worse proposal mean scores than men in almost every observing semester, a difference that is statistically significant in semesters 2013A, 2014A, and 2016B for both the sample as a whole as well as for the subsample of non-faculty PIs. 

Given the statistically significant t-tests for subsets of the data illustrated in Figures~\ref{fig:telescopes},~\ref{fig:panel}~and~\ref{fig:semester}, are there other predictors of mean proposal scores besides PI sex? As described in Section~\ref{sec:methods}, we address this question by carrying out a multivariate regression analysis which assesses the effect of PI sex on proposal scores by controlling for other covariates: telescope, panel, semester, and, for faculty PIs, PhD year. Table~\ref{tab:multivariate} shows the results. For the sample as a whole, we find that PI sex is the only significant predictor of proposal scores, net of other covariates. As already mentioned above, we find no significant predictors of proposal scores at all (even PI gender) for the faculty PI subsample. However, both PI sex and panel are significant predictors of proposal score for non-faculty PIs: women in that subsample are significantly less likely to obtain better proposal scores than men, and non-faculty PI proposals evaluated by the Extragalactic panel are significantly less likely to obtain better proposal scores than non-faculty PI proposals evaluated by the Galactic panel.

\section{DISCUSSION AND CONCLUSIONS}
\label{sec:discuss}

Using data from 10 recent observing semesters provided to us by NRC, we have investigated whether or not mean proposal scores assigned by CanTAC for access to Canadian CFHT and Gemini time differ significantly for men and women. We use t-tests (Table~\ref{tab:ttests}) as well as bivariate regressions (Table~\ref{tab:bivariate}) to examine differences in mean proposal scores for men and women in the sample as a whole, and for subsamples containing only faculty PIs and only non-faculty PIs. We also searched for systematic trends in proposal scores in relation to the telescope requested, the scientific panel that reviewed the proposal, the observing semester, and, for faculty PIs, the year in which their PhD was awarded (Tables~\ref{tab:ttests}~and~\ref{tab:multivariate}).  We find that, for the sample as a whole, the mean proposal scores for women differ significantly from those of men, in the sense that men are more likely to obtain better proposal scores than women (Figure~\ref{fig:alldata}). 

The sex-disaggregated systematics that we find in the sample as a whole are also present in the non-faculty PI subsample: specifically, our t-tests show that women are significantly less likely to receive comparable proposal scores to men for access to Gemini (Figure~\ref{fig:telescopes}), for proposals reviewed by the Extragalactic panel (Figure~\ref{fig:panel}), and for the individual observing semesters 2013A, 2014A, and 2016B (Figure~\ref{fig:semester}). We also find that women in the faculty PI subsample who were awarded PhDs from 1990--1999 have significantly worse mean proposal scores than men awarded PhDs in the same period (Figure~\ref{fig:phdyear}). We note that even for subsets of the data where the difference between mean proposal scores for men and women is not statistically significant, those for women are almost always worse than those for men (Table~\ref{tab:ttests}, Figure~\ref{fig:semester}).   Our multivariate regressions imply that, for the sample as a whole, PI sex is the only significant predictor of proposal scores, while both PI sex and scientific panel are  significant predictors of non-faculty PI proposal scores (Table~\ref{tab:multivariate}). 

The results presented here add to the growing list of studies that have uncovered sex-disaggregated systematics in the outcomes of peer review processes designed to evaluate scientific and technical merit, within both the relatively narrow topic of telescope time allocation \cite{Reid14,Patat16,Lonsdale16} and within the broader context of review outcomes\cite{Wenneras97,Conley12,Moss12,West13,Caplar16,King16} and gender statistics \cite{UN12} for women in academia.  In the present study, the statistically significant differences that we found in the non-faculty PI subsample often mirror those found in the sample as a whole. This suggests that the trends in the former are significant drivers of those in the latter, although more research is needed to verify this hypothesis.  More work is also needed to understand scientific panel as a significant predictor of mean proposal scores for non-faculty PIs in our multivariate regressions. We note that various sub-disciplines within Canadian astronomy consistently fall within the purview of either the Extragalactic or Galactic panel, and therefore a difference in the climate for young researchers within these sub-disciplines may explain this trend.

A thorough examination of the underlying causes for the significant differences that we find in the sex-disaggregated CanTAC proposal scores, and how they may inform the broader issue of gender systematics in academia \cite{UN12}, is beyond the scope of this work. The hypothesis advanced by Patat \cite{Patat16} that men obtain better proposal scores than women because they have a higher average seniority is unlikely to apply to our study because some of the most significant systematics uncovered are within the non-faculty PI subsample, where there is little difference in seniority among applicants.  We find it more plausible that implicit social cognition \cite{Greenwald95,Carney10,Nosek11}, or processes in human cognition outside of conscious control that influence judgement and action, is at work. Studies show that implicit cognition plays a role in stereotyping by sex \cite{Stroessner96,Pratto91}, and research is ongoing using tools like the Implicit Association Test\footnote{https://implicit.harvard.edu/implicit/} \cite{Greenwald02,Nosek07}. 

An important challenge for telescope operations is to develop policies that mitigate systematics other than scientific and technical merit in proposal reviews. We note that if implicit social cognition is a factor in the sex-disaggregated statistics reported here and elsewhere, then altering the demographics of review panels is unlikely to produce different outcomes since numerous studies have shown that both men and women consistently rank work done by men more highly than identical work done by women \cite{Wenneras97,Moss12}. This issue was also examined in the particular context of telescope proposal reviews by Lonsdale et al.\ \cite{Lonsdale16}, who found no correlation between proposal ranking trends and the fraction of women on review panels.  
 
A potential avenue for mitigating bias in sex-disaggregated statistics that stems from implicit cognition is instead to obscure the identity of the proposer(s) from the reviewers. Similar ``blind" evaluation strategies have produced dramatic demographic shifts in at least some areas, ranging from the fraction of ecology journal articles written by women \cite{Budden08} to the fraction of women who successfully audition for symphony orchestras \cite{Goldin00}. Starting in semester 2017B, CanTAC has adopted a similar approach to that for HST proposal reviews\footnote{www.stsci.edu/hst/proposing/panel} by obscuring PI and co-investigator names in the documents sent to CanTAC members and external referees. While it is still too early to tell if this strategy has changed CanTAC's sex-disaggregated statistics relative to those reported here, we expect that $\sim$5 semesters worth of data would be sufficient for a quantitative first look. We recommend that the impact of any mitigation strategies adopted on the sex-disaggregated systematics in CanTAC data be quantitatively assessed using robust statistical techniques such as those adopted here.

\acknowledgments 
We thank NRC for funding this study and granting permission to publish the results. We also thank Masato Onodera for pointing out inconsistencies between the figures and tables in an earlier version of this contribution.

\newpage
\bibliography{report} 

\begin{thebibliography}{10}

\bibitem{Reid14}
{Reid}, I.~N., ``{Gender-Correlated Systematics in HST Proposal Selection},''
  {\em PASP}~{\bf 126},  923 (Oct. 2014).

\bibitem{Patat16}
{Patat}, F., ``{Gender Systematics in Telescope Time Allocation at ESO},'' {\em
  The Messenger}~{\bf 165},  2--9 (Sept. 2016).

\bibitem{Lonsdale16}
{Lonsdale}, C.~J., {Schwab}, F.~R., and {Hunt}, G., ``{Gender-Related
  Systematics in the NRAO and ALMA Proposal Review Processes},'' {\em ArXiv
  e-prints}  (Nov. 2016).

\bibitem{Wenneras97}
{Wenner\aa s}, C. and Wold, A., ``{Nepotism and sexism in peer-review},'' {\em
  Nature}~{\bf 387},  341--343 (1997).

\bibitem{Conley12}
{Conley}, D. and {Stadmark}, J., ``{A call to commission more women writers},''
  {\em Nature}~{\bf 488},  590--591 (Aug. 2012).

\bibitem{Moss12}
{Moss-Racusin}, C., {Dovidio}, J., {Brescoll}, V., {Graham}, M., and
  {Handelsman}, J., ``{Science faculty's subtle gender biases favor male
  students},'' {\em Proceedings of the National Academy of Sciences}~{\bf 109},
   1474 (2012).

\bibitem{West13}
{West}, J.~D., {Jacquet}, J., {King}, M.~M., {Correll}, S.~J., and {Bergstrom},
  C.~T., ``{Gender Systematics in Telescope Time Allocation at ESO},'' {\em The
  Messenger}~{\bf 165},  2--9 (Sept. 2016).

\bibitem{Caplar16}
{Caplar}, N., {Tachella}, S., and {Birrer}, S., ``{Quantitative evaluation of
  gender bias in astronomical publications from citation counts},'' {\em
  Nature}~{\bf 1} (May 2017).

\bibitem{King16}
{King}, M.~M., {Bergstrom}, T.~C., {Shelley}, C.~J., {Jacquet}, J., and {West},
  J.~D., ``{Men set their own cites high: gender and self-citation across
  fields over time},'' {\em ArXiv e-prints}  (2016).

\bibitem{UN12}
Zheng, S.,  [{\em Gender Systematics Manual of the United Nations Statistical
  Commission}{\nolinebreak\hspace{0.1em}]}, United Nations
  (https://unstats.un.org/unsd/genderstatmanual/), New York (2015).

\bibitem{Greenwald95}
{Greenwald}, A.~G. and {Banaji}, M.~R., ``{Implicit social cognition:
  Attitudes, self-esteem, and stereotypes},'' {\em Psychological Review}~{\bf
  102},  4--27 (Jan. 1995).

\bibitem{Carney10}
{Carney}, D., {Krieger}, N., and {Banaji}, M.~R., ``{Implicit Measures Reveal
  Evidence of Personal Discrimination},'' {\em Self and Identity}  (Nov. 2010).

\bibitem{Nosek11}
{Nosek}, B.~A., {Hawkins}, C.~B., and {Frazier}, R.~S., ``{Implicit social
  cognition: From measures to mechanisms},'' {\em Trends Cog. Sci}  (Apr.
  2011).

\bibitem{Stroessner96}
{Stroessner}, S.~J., ``{Social categorization by race or sex: Effects of
  perceived non-normalcy on response times},'' {\em Social Cognition}~{\bf 14},
   247--276 (1996).

\bibitem{Pratto91}
{Pratto}, F. and {Bargh}, J.~A., ``{Stereotyping based on apparently
  individuating information: Trait and global components of sex stereotypes
  under attention overload},'' {\em Journal of Experimental Social
  Psychology}~{\bf 27},  26--47 (1991).

\bibitem{Greenwald02}
{Greenwald}, A.~G., {Banaji}, M.~R., Rudman, L.~A., Nosek, B.~A., and Mellot,
  D.~S., ``{A unified theory of implicit attitudes, beliefs, self-esteem and
  self-concept},'' {\em Psychological Review}~{\bf 109},  3--25 (1995).

\bibitem{Nosek07}
{Nosek}, B.~A., {Greenwald}, A.~G., and {Banaji}, M.~R., ``The implicit
  association test at age 7: A methodological and conceptual review,'' in [{\em
  Social Psychology and the Unconscious: The Automaticity of Higher Mental
  Processes}{\nolinebreak\hspace{0.1em}]},  Bargh, J.~A., ed.,  265--292,
  Psychology Press, New York (2007).

\bibitem{Budden08}
{Budden}, D., {Tregenza}, T., {Aarssen}, S., {Leimu}, R., and Lortie, C.~J.,
  ``{Double-blind review favours increased representation of female authors},''
  {\em Trends in Ecology and Evolution}~{\bf 23},  4--6 (Jan. 2008).

\bibitem{Goldin00}
{Goldin}, C. and {Rouse}, C., ``{Orchestrating Impartiality: The Impact Of
  `Blind' Auditions On Female Musicians},'' {\em American Economic Review}~{\bf
  4},  715--741 (Sept. 2000).

\end{thebibliography}
\bibliographystyle{spiebib} 

\end{document}